\documentclass[aps,prd,a4paper,10pt,superscriptaddress,preprintnumbers]{revtex4}

\usepackage[dvipdfm]{graphicx,color}
\usepackage{amsmath,amssymb,bm}
\usepackage{braket}
\usepackage{geometry}
\usepackage{enumerate}
\usepackage{url}
\usepackage{mathrsfs}
\usepackage{color}
\usepackage{booktabs} 
\usepackage{array} 
\usepackage{paralist} 
\usepackage{verbatim} 

\usepackage{layout}
\allowdisplaybreaks[4]

\begin{document}
\title{Improving parameter estimation accuracy with torsion-bar antennas}
\author{Kazunari Eda}
\email{eda@resceu.s.u-tokyo.ac.jp}
\affiliation{
Department of Physics, 
Graduate School of Science, 
University of Tokyo, Tokyo 113-0033, Japan
}      
\affiliation{
Research center for the early universe (RESCEU), 
Graduate School of Science,   
University of Tokyo, Tokyo 113-0033, Japan
}     
\author{Ayaka Shoda}
\email{shoda@granite.phys.s.u-tokyo.ac.jp}
\affiliation{
Department of Physics, 
Graduate School of Science, 
University of Tokyo, Tokyo 113-0033, Japan
} 
\author{Yousuke Itoh}
\email{yousuke_itoh@resceu.s.u-tokyo.ac.jp}
\affiliation{ 
Research center for the early universe (RESCEU),   
Graduate School of Science, 
University of Tokyo, Tokyo 113-0033, Japan
}   
\author{Masaki Ando}
\email{ando@phys.s.u-tokyo.ac.jp}
\affiliation{
Department of Physics, 
Graduate School of Science, 
University of Tokyo, Tokyo 113-0033, Japan
}
\affiliation{ 
Research center for the early universe (RESCEU),   
Graduate School of Science, 
University of Tokyo, Tokyo 113-0033, Japan
}  
\affiliation{
Gravitational Wave Project Office, Optical and Infrared Astronomy Division, 
National Astronomical Observatory, Osawa 2-21-1, Mitaka, 
Tokyo 181-8588, Japan
}

\begin{abstract}
We propose a new antenna configuration of a torsion-bar antenna (TOBA) and study its performance. 
A TOBA is a novel type of an antenna for low-frequency gravitational waves (GWs) 
which consists of two bar-shaped orthogonal test masses. 
Previously only the rotation of the bars on the horizontal plane had been considered as output signals. 
In this paper, we introduce a new antenna configuration for a TOBA 
to incorporate two additional outputs by measuring the rotation of the bars on the vertical planes. 
Such a triple-output TOBA can be regarded 
as a network of three coincident but misaligned interferometric detectors. 
We investigate its event detection rate and its parameter accuracies using Fisher analysis. 
We find that since the triple-output TOBA can discriminate 
two polarization modes of a short-duration GW signal even with a single antenna 
thanks to  having three independent outputs, 
it improves the detection rate and the accuracies of waveform parameters drastically. 
\end{abstract} 
 
\preprint{RESCEU-35/14}
\maketitle   
 
\section{Introduction}\label{Sec:Intro}
A torsion-bar Antenna (TOBA) is a gravitational-wave (GW) detector with two bar-shaped test masses 
which rotate differentially by the tidal force of GWs \cite{Ando:2010zz, Ishidoshiro:2011ii, Shoda:2013oya, Nakamura:2014wia}.
The main feature of the TOBA is that it has good sensitivity at low frequencies around 1 Hz even on the ground
because of its low resonant frequency of the test masses in the rotational degrees of freedom. 
Other ground-based detectors such as LIGO \cite{Abbott:2007kv}, Virgo \cite{Accadia:2012zzb} and KAGRA \cite{Aso:2013eba} 
are sensitive to GWs only above about 10 Hz.
The test masses of interferometers do not behave as free masses and the seismic noise is not reduced below 10 Hz 
since the resonant frequency of the pendulum mode is higher than that of torsional mode.
Though space-borne GW detectors such as evolved Laser Interferometer Space Antenna (eLISA) \cite{AmaroSeoane:2012je} 
and DECi-hertz Interferometer Gravitational wave Observatory (DECIGO) \cite{Kawamura:2011zz} are sensitive at low frequencies, 
such space missions need several decades to be launched.
TOBA is the possible way to detect low-frequency GWs with less time and fewer costs.

The main observational targets of a TOBA are GWs from binaries of intermediate mass black holes (IMBHs) and pulsars.
The individual IMBHs have not been directly detected yet, 
though there are several evidences for the existence of IMBHs such as ultraluminous X-ray sources \cite{Miller:2003sc}.
The direct detection of IMBHs is important to understand the evolution of SMBHs and galaxies.
Long period pulsars whose rotational period are of the order of 1-10 seconds such as PSR J2144-3933 \cite{Young:1999} 
are also interesting targets.
It is necessary to estimate binary parameters or pulsar parameters such as their masses, spin rate and direction, from GW signals 
for revealing the nature of the IMBHs and the neutron stars if the signal has been detected.
However, it is difficult to measure the GW waveform parameters and the direction of the source with a single-detector observation  
when the observation time is not long enough to use the effect of the spin of the Earth 
because the conventional GW detectors do not have good directivity. 
Estimating the direction of the GW sources enables us to perform 
complementary observations with electro-magnetic waves which would provide us more detailed information.

In this paper, we propose a new observational configuration 
for TOBA to improve the accuracy of the parameter estimation, the angular resolution and event detection rates without changing the setup drastically.
We can obtain three independent signals by monitoring not only horizontal rotation of the test mass, but also its vertical rotations.
Since these three signals have different responses to GWs in such a way that 
they compensate their blind directions each other, the detection volume is enlarged and the detection rate is improved.
Furthermore, the waveform parameters can be determined with a single TOBA even though the observation time is shorter than a day.
It will enable us to derive more information about GW sources with less detectors.

The rest of this paper is composed as follows.  
We present the details of our antenna configuration and 
the antenna pattern functions in section \ref{Sec:Obs_method}.
In section \ref{Sec:Para_resol}, we show the accuracy of the parameter estimation evaluated by the Fisher analysis 
for both monochromatic sources and binary coalescences.
Summary and conclusions are presented in section \ref{Sec:Conc}. 

\section{New antenna configuration}\label{Sec:Obs_method}
\subsection{Setup}
\begin{figure}[tbh]
\centering
\includegraphics[width=18pc]{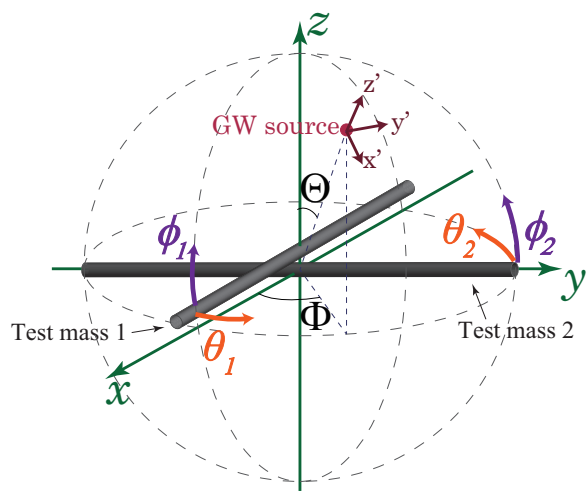}
\caption{\label{fig:bar}Schematic picture of TOBA in the proper antenna frame. 
The test masses rotate in the horizontal plane ($\theta_1$ and $\theta_2$) 
and in the vertical planes ($\phi_1$ and $\phi_2$) due to the tidal force of GWs.}
\end{figure}

\begin{figure}[tbh]
\centering
\includegraphics[width=8cm,clip]{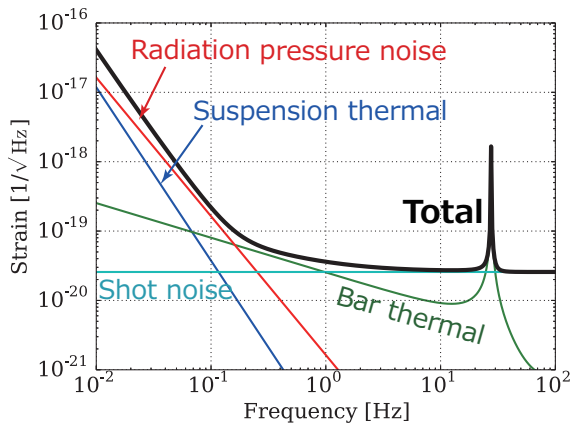}
\caption{\label{fig:sens}The estimated sensitivity curve of 10-m scaled TOBA.}
\end{figure}

TOBA is a gravitational-wave antenna with two orthogonal bar-shaped test masses. 
In the presence of GW signal $h_{ij}$, the bars are subjected to the tidal force of GWs and rotate. 
Then they work as test masses for GW detection above the resonant frequency. 
The horizontal angular motion of the test mass 1 shown in Fig. \ref{fig:bar}, $\theta_1$ obeys the equation of motion \cite{Ando:2010zz}:
\begin{equation} 
I \ddot{\theta_1}\left(t\right) + \gamma_{\theta} \dot{\theta_1}\left(t\right) + \kappa_{\theta} \theta_1\left(t\right) = \frac{1}{4} \ddot{h}_{ij}q_{\theta_1}^{ij},
\end{equation} 
where $I, \gamma, \kappa$ and $q_{\theta_1}^{ij}$ are the moment of inertia, dissipation coefficient and spring constant in the rotational degree of freedom, 
and the dynamic quadrupole moment of the horizontal rotational mode  \cite{Hirakawa:1976}, respectively. 
Then, TOBA has sensitivity to GWs above the resonant frequency $f_{\theta 0} = \sqrt{\kappa_{\theta}/I}/2\pi$.
When we assume that the each bar is suspended by two wires, the estimated example sensitivity curve and its parameters are shown in Fig. \ref{fig:sens} and Table \ref{tbl:params}.
Note that the spring constant $\kappa$ is derived by $\kappa = mga^2/l$, where $g$ is the gravity acceleration, $a$ is the distance between the two wires, and $l$ is the length of the suspension wires.  
\begin{table}[h]\begin{center}  
\begin{tabular}{lll}
\hline \hline
Test mass & Material & Alminum \\ 
&Length of the bar & 10 [m] \\ 
&Diameter of the bar & 0.6 [m] \\
&Mass ($m$) & 7400 [kg] \\ 
&Moment of inertia ($I$) & $6.4 \times 10^4$ [N m s$^2$] \\
&Loss angle & $10^{-7}$ \\ 
\hline
Suspension & Distance between the two suspension points ($a$) & 5 [cm] \\
& Length of the suspension wires ($l$) & 3 [m] \\
& Dissipation coefficient ($\gamma_{\theta}$) & $1.0\times10^{-7}$ \\ 
\hline
Fabry-Perot laser && \\
interferometric sensor & Wave length & 1064 [nm] \\
&Power & 10 [W] \\ 
&Finesse & 100 \\ \hline \hline
\end{tabular}
\caption{\label{tbl:params}Parameters of TOBA used for the estimated sensitivity curve shown in Fig. \ref{fig:sens}}\end{center}
\end{table}

Meanwhile, the angular motion in the vertical plane $\phi_1$ obeys the equation of motion
\begin{equation}
I \ddot{\phi_1}(t) + \gamma_{\phi} \dot{\phi_1}(t) + \kappa_{\phi} \phi_1(t) = \frac{1}{4} \ddot{h}_{ij}q_{\phi_1}^{ij}.
\end{equation}
The notation is basically the same as above.
It means that the GW signals also appear in $\phi_1$.
 
In this case, $\kappa_{\phi}$ is written as
\begin{equation}
\kappa_{\phi} = mgd,
\end{equation}
where $d$ is the vertical distance between the center of mass of the test mass and the suspension point.
Therefore, the resonant frequency $f_{\phi 0} = \sqrt{\kappa_{\phi}/I}/2\pi$ will be below 10 mHz when the center of mass is adjusted so that $d \leq 3 {\rm mm}$.

The estimated sensitivity obtained by monitoring $\phi$ would be basically the same as $\theta$ except the suspension thermal noise when the configuration of the sensors are the same.
The suspension thermal noise would be larger by a factor of $f_{\phi0}/f_{\theta0}\sqrt{Q_{\theta}/Q_{\phi}}$, where $Q_{i}$ is the quality factor of the fibers in the direction of $i$.

The same equation of motion is applied for the test mass 2.
Therefore, considering $\theta_2 = - \theta_1$ \cite{Ando:2010zz}, 
TOBA can obtain three independent outputs, 
$s_{\text{I}} = \theta_1 - \theta_2, s_{\text{II}} = \phi_1$, and $s_{\text{III}} = \phi_2$, without changing the setup drastically.
All we additionally need is sensors to monitor $\phi_1$ and $\phi_2$. 
In the following, the output obtained from 
$ \theta_1 - \theta_2, \phi_1$, and $\phi_2$ 
are referred to as the output I, II and III, respectively.

\subsection{Antenna response}
The response of the antenna to the incident GW depends on its relative position and  orientation to the GW source. 
Such geometrical information is encoded in 
antenna pattern functions  $F_+$ and $F_{\times}$ which correspond to the two independent GW polarizations. 
Using $F_+$ and $F_{\times}$, the GW signal $h\left(t\right)$ in the antenna output is expressed by 
\begin{align}
 h\left(t\right) = F_+ \left(t\right) h_+\left(t\right) + F_{\times} \left( t\right) h_{\times}\left(t\right).  \label{Eq:GW_signal}
\end{align}  
Let us introduce two coordinate systems, a proper antenna frame $\left(x,y,z\right)$ such that 
the two bars are alined with the $x$-axis and the $y$-axis as illustrated in Fig. \ref{fig:bar} and 
a wave-coming frame $\left(x',y',z'\right)$ such that the direction of GW propagation is along the $z'$-axis. 
The GW waveform $\boldsymbol{h}\left(t\right)$ in the proper antenna frame 
can be related to the waveform $\boldsymbol{h}'\left(t\right)$ in the wave-coming frame by 
\begin{align}
 \boldsymbol{h}\left(t\right) = M\left(t\right) \boldsymbol{h}'\left(t\right) M\left(t\right)^{\text{T}} \label{Eq:GWresponse_Mmatrix}
\end{align} 
where $M$ is the 3-dimensional transformation matrix and $M^{\text{T}}$ denotes the transpose of the matrix $M$. 

As discussed in the section \ref{Sec:Obs_method}, the two orthogonal bars are 
twisted differentially by the tidal force from the incoming GW. 
When the tidal force produces the small rotation of the bar on the $x$-axis toward the $y$-direction, 
the resulting GW signal is expressed by  $h_{ij} n_x^i n_y^j$. 
Similarly, the GW signals $h_{\text{I}} , h_{\text{II}} $ and $h_{\text{III}} $ of the three outputs 
obtained from $\theta_1 - \theta_2, \phi_1$ and $\phi_2$ in Fig. \ref{fig:bar}  are expressed by 
\begin{subequations}
\begin{align}
 h_{\text{I}}    &= \dfrac{1}{2} \left( n_x^i n_y^j + n_y^i n_x^j \right) h_{ij} = n_x^i n_y^j h_{ij}, \label{Eq:def_hI}\\
 h_{\text{II}}   &= \dfrac{1}{2} n_x^i n_z^j h_{ij}, \label{Eq:def_hII}\\
 h_{\text{III}}  &= \dfrac{1}{2} n_y^i n_z^j h_{ij} \label{Eq:def_hIII}
\end{align}
\end{subequations}
where  $\boldsymbol{n}_x, \boldsymbol{n}_y$ and $\boldsymbol{n}_z$ 
are the unit vectors pointing toward the $x,y$ and $z$-directions respectively  in the proper antenna frame. 
$h_{\text{I}} , h_{\text{II}} $ and $h_{\text{III}} $ can be decomposed into 
the sum of the two independent polarization $h_+$ and $h_\times$ as in Eq.(\ref{Eq:GW_signal}). 
Combining Eqs.(\ref{Eq:GWresponse_Mmatrix})-(\ref{Eq:def_hIII}), 
we can find the explicit expressions for the antenna pattern functions. 

When the signal duration is so short that  
the relative motion of the antenna with respect to the sources is negligible, 
$F_+$ and $F_\times$ can be regarded as constants. 
In this case, the pattern functions are given by 
\begin{subequations}
\begin{align}
 F^+_{\text{I}}       &= -\dfrac{1+\cos^2\Theta}{2}\sin 2\Phi \cos\Psi + \cos\Theta \cos2\Phi\sin2\Psi, \label{Eq:F1p_short}\\
 F^\times_{\text{I}}  &= \cos\Theta \cos2\Phi\cos2\Psi + \dfrac{1+\cos^2\Theta}{2}\sin 2\Phi\sin2\Psi, \label{Eq:F1c_short}\\
 F^+_{\text{II}}      &= \dfrac{1}{4}\sin2\Theta\sin\Phi\cos2\Psi - \dfrac{1}{2}\sin\Theta\cos\Phi\sin2\Psi,\\
 F^\times_{\text{II}} &= -\dfrac{1}{2}\sin\Theta\cos\Phi\cos2\Psi + \dfrac{1}{4}\sin2\Theta\sin\Phi\sin2\Psi,\\
 F^+_{\text{III}}     &= \dfrac{1}{4} \sin 2\Theta\cos\Phi\cos2\Psi + \dfrac{1}{2}\sin\Theta\sin\Phi\sin2\Psi,\\
 F^\times_{\text{III}}&= \dfrac{1}{2}\sin\Theta\sin\Phi\cos2\Psi - \dfrac{1}{4} \sin 2\Theta\cos\Phi\sin2\Psi \label{Eq:F3c_short}
\end{align}
\end{subequations}
where $\Psi$ is the GW polarization angle. 
The direction of the incoming GW is assumed to be 
$\boldsymbol{n} = -\left(\sin\Theta\sin\Phi, \sin\Theta\cos\Phi,\cos\Theta\right)$
in the proper antenna frame.  
The antenna responses of the single-output TOBA 
shown in Eqs.(\ref{Eq:F1p_short}) and (\ref{Eq:F1c_short}) 
are basically equivalent to that of an interferometric detector.

For long-duration signals, the antenna and the source 
cannot be treated as at rest with respect to each other 
because of the Earth's rotation and revolution.  
So $\Theta$ and $\Phi$ in Eqs. (\ref{Eq:F1p_short})-(\ref{Eq:F3c_short}) are not constant any longer. 
These relative motions induce the amplitude-modulation and phase-modulation of the signal. 
To take into account these effects, we follow the formulation 
presented by Jaranowski, Krolak and Schutz \cite{Bonazzola:1995rb, Jaranowski:1998qm} 
and obtain the following expressions for the antenna pattern functions. 
\begin{subequations}
\begin{align}  
 &F^+_N\left(t\right) = a_N\left(t\right) \cos 2\psi + b_N \left(t\right) \sin 2\psi, \label{Eq:F+}\\
 &F^\times_N\left(t\right) = b_N \left(t\right) \cos 2\psi -  a_N\left(t\right) \sin 2\psi, \label{Eq:Fx}\\
 &N = \text{I},\text{II}, \text{III}. \nonumber
\end{align}
\end{subequations}
The modulation amplitudes $a_N\left(t\right)$ and $b_N\left(t\right)$ are given by 
\begin{subequations}
\begin{align}
 a_{\text{I}}  \left(t\right)
   &= \dfrac{3}{4}\cos 2\gamma \cos^2 \delta \cos^2 \lambda \nonumber\\
   &\hspace{1em} + \dfrac{1}{16}\cos 2\gamma \left( 3-\cos 2\delta \right)\left( 3-\cos 2\lambda \right)\cos \left[2\left(\alpha - \phi_r-\Omega_r t \right) \right] \nonumber \\
   &\hspace{1em} + \dfrac{1}{4} \cos 2\gamma\sin 2\lambda  \sin 2\delta \cos\left(\alpha - \phi_r-\Omega_r t\right) \nonumber\\
   &\hspace{1em} + \dfrac{1}{2}\sin 2\gamma \cos \lambda \sin 2\delta \sin \left( \alpha - \phi_r-\Omega_r t\right)  \nonumber \\
   &\hspace{1em} + \dfrac{1}{4}\sin 2\gamma \left( 3-\cos 2\delta \right) \sin \lambda \sin \left[ 2\left( \alpha - \phi_r-\Omega_r t\right) \right], \\
  b_{\text{I}}  \left(t\right)
   &= - \sin 2\gamma\cos \delta \cos \lambda \cos \left(\alpha - \phi_r-\Omega_r t\right) \nonumber\\
   &\hspace{1em} - \sin 2\gamma \sin \delta \sin \lambda \cos \left[2\left(\alpha - \phi_r-\Omega_r t\right) \right] \nonumber \\
   &\hspace{1em} + \dfrac{1}{2} \cos 2\gamma \cos \delta \sin 2\lambda \sin \left(\alpha - \phi_r-\Omega_r t\right) \nonumber\\
   &\hspace{1em} + \dfrac{1}{4} \cos 2\gamma \left( 3-\cos 2\lambda \right) \sin \delta \sin \left[2\left(\alpha - \phi_r-\Omega_r t\right)\right],\\
  a_{\text{II}}  \left(t\right) 
   &= \dfrac{1}{2}\sin \delta \cos \delta \sin \left(\gamma + \dfrac{\pi}{4}\right) \sin \lambda \sin \left(\alpha - \phi_r-\Omega_r t\right)  \nonumber\\
   &\hspace{1em} - \dfrac{1}{4} \cos\left(\gamma + \dfrac{\pi}{4}\right)\cos 2\lambda \cos\left(\alpha-\phi_r-\Omega_r t\right)\sin 2\delta  \nonumber\\
   &\hspace{1em} + \dfrac{1}{16} \cos\left(\gamma + \dfrac{\pi}{4}\right)\cos 2\delta \left(3+ \cos\left[2\left(\alpha-\phi_r-\Omega_r t\right)\right] \right) \sin 2\lambda  \nonumber\\
   &\hspace{1em} + \dfrac{3}{8}\cos\left(\gamma + \dfrac{\pi}{4}\right)\sin 2\lambda \sin^2\left(\alpha-\phi_r-\Omega_r t\right)  \nonumber\\
   &\hspace{1em} - \frac{1}{8} \left(3-\cos\delta\right)\cos\lambda \sin\left(\gamma + \dfrac{\pi}{4}\right) \sin\left[2\left(\alpha-\phi_r-\Omega_r t\right)\right], \\  
   b_{\text{II}} \left(t\right)
   &=-\dfrac{1}{2}\cos\delta \cos\left(\alpha- \phi_r-\Omega_r t\right)\sin\left(\gamma + \dfrac{\pi}{4}\right)\sin \lambda \nonumber\\
   &\hspace{1em} - \dfrac{1}{2}\cos\delta \cos\left(\gamma + \dfrac{\pi}{4}\right)\cos 2\lambda\sin\left(\alpha-\phi_r-\Omega_r t\right) \nonumber\\
   &\hspace{1em} + \dfrac{1}{2}\sin\delta \cos\lambda\cos\left[2\left(\alpha-\phi_r-\Omega_r t\right)\right]\sin\left(\gamma + \dfrac{\pi}{4}\right) \nonumber\\
   &\hspace{1em} - \dfrac{1}{4}\sin\delta\cos\left(\gamma + \dfrac{\pi}{4}\right)\sin 2\lambda \sin\left[2\left(\alpha-\phi_r-\Omega_r t\right)\right], \\
  a_{\text{III}} \left(t\right)
   &=-\dfrac{3}{8}\cos^2\delta \sin\left(\gamma + \dfrac{\pi}{4}\right)\sin 2\lambda \nonumber\\
   &\hspace{1em} + \dfrac{1}{4}\sin 2\delta \cos 2\lambda \cos \left(\alpha - \phi_r-\Omega_r t \right) \sin \left(\gamma + \dfrac{\pi}{4}\right) \nonumber\\
   &\hspace{1em} + \dfrac{1}{4}\sin 2\delta \sin \lambda  \sin \left(\alpha - \phi_r-\Omega_r t \right) \cos\left(\gamma + \dfrac{\pi}{4}\right) \nonumber\\
   &\hspace{1em} + \dfrac{1}{8}\left( 3-\cos 2\delta \right) \cos \lambda \cos \left[ 2\left( \alpha - \phi_r-\Omega_r t \right) \right] \sin \left(\gamma + \dfrac{\pi}{4}\right) \sin \lambda \nonumber\\
   &\hspace{1em} -\dfrac{1}{8} \left(3-\cos 2\delta \right) \cos \lambda \cos \left(\gamma + \dfrac{\pi}{4}\right) \sin \left[ 2\left(\alpha - \phi_r-\Omega_r t \right) \right], \\
  b_{\text{III}} \left(t\right) 
   &= - \dfrac{1}{2}\cos\left(\gamma+\dfrac{\pi}{4}\right)\cos \delta \cos\left(\alpha-\phi_r-\Omega_r t\right)\sin \lambda  \nonumber\\
   &\hspace{1em} + \dfrac{1}{2}\cos\delta \cos 2\lambda \sin\left(\gamma+\dfrac{\pi}{4}\right)\sin\left(\alpha-\phi_r-\Omega_r t\right)   \nonumber\\
   &\hspace{1em} + \dfrac{1}{2}\cos\left(\gamma +\dfrac{\pi}{4}\right)\cos \lambda \cos\left[2\left(\alpha-\phi_r-\Omega_r t\right)\right]\sin \delta   \nonumber\\
   &\hspace{1em} + \dfrac{1}{2}\cos\delta\cos 2\lambda \sin\left(\gamma+\dfrac{\pi}{4}\right)\sin\left(\alpha-\phi_r-\Omega_r t\right)  \nonumber\\
   &\hspace{1em} + \dfrac{1}{4}\sin\left(\gamma+\dfrac{\pi}{4}\right) \sin\delta \sin 2\lambda \sin\left[2\left(\alpha-\phi_r-\Omega_r t\right)\right].
\end{align}
\end{subequations}
where $\alpha$ and $\delta$ specify the sky position of the GW sources and are called right ascension and declination, respectively. 
$\lambda, \gamma, \Omega_r$ and $\phi_r$ are the latitude of the antenna position, the angle between the local East direction 
and the bisector of the antenna, the angular speed of the Earth, 
and the initial phase which defines the antenna position on the Earth at $t=0$, respectively.
\subsection{Detection volume}
\begin{figure} 
\centering 
\includegraphics[width=15cm,clip]{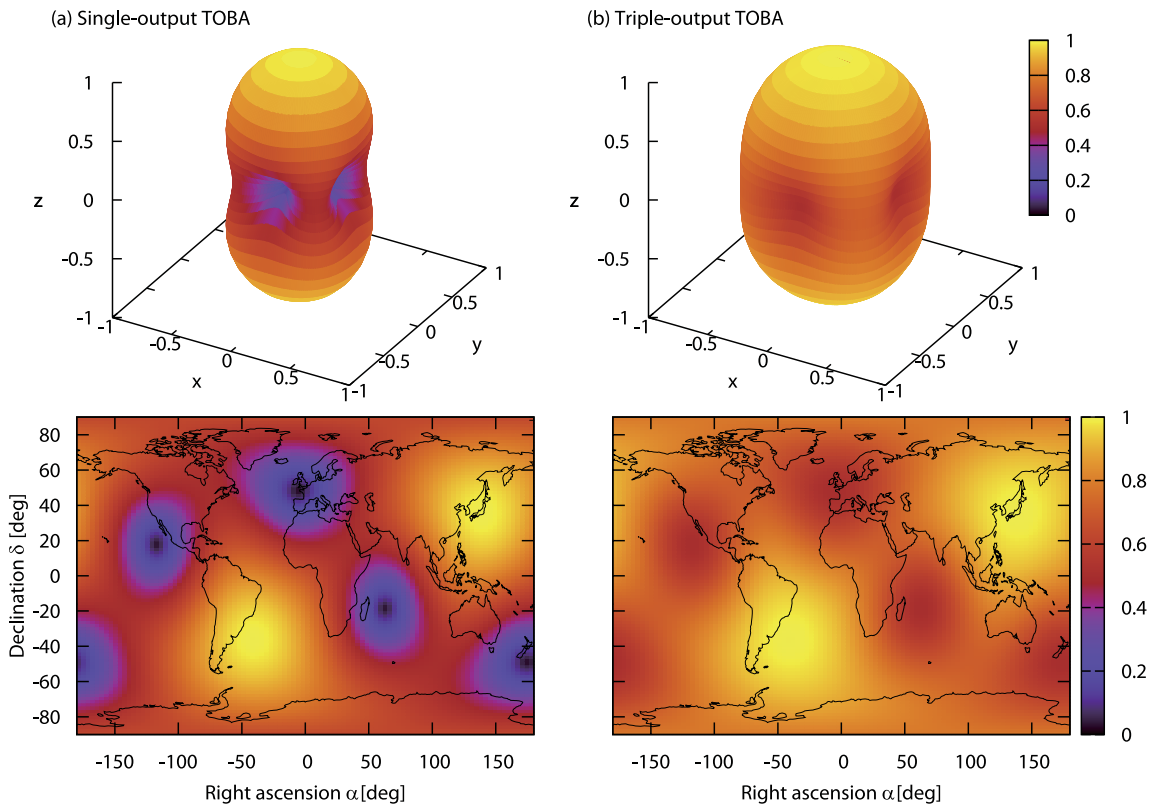}
\caption{
Square root of the antenna power patterns $\sqrt{P_N}$ for 
the single-output TOBA (left panel) and 
the triple-output TOBA (right panel). 
The antennas are assumed to be located at the TAMA300's site in Japan 
($ 35^{\circ} \ 40' \ 36''$ N, $139^{\circ} \ 32' \ 10''$ E). 
} 
\label{Fig:AntennaPower}
\end{figure}

The method presented in the section \ref{Sec:Obs_method} enhances the detection volume
\begin{equation}
V_N = \int {\rm d}\Omega \int_{0}^{R_N \left(\alpha, \delta\right)}r^2 {\rm d}r,
\end{equation}
which represents the region enclosed by its reach in any direction \cite{Schutz:2011tw} with 
\begin{equation}
R_N\left(\alpha, \delta\right) = \frac{D}{\rho_{\rm min}}\left[ P_N\left(\alpha, \delta\right) \right]^{1/2}. 
\end{equation}
Here $D$ is the distance to which the antenna can observe GWs by a single signal with a unit signal-to-noise ratio (SNR), 
$\rho_{\text{min}}$ is the minimum value of the SNR above which we can regard that the GW signal is detected, and 
$P_N \left(\alpha, \delta\right) = \sum_{N}\left(F_{N}^{+}{}^2 + F_{N}^{\times}{}^2 \right)$ 
is the antenna pattern power function depicted in Fig. \ref{Fig:AntennaPower}. 
In our case, the summation is taken over the three outputs I, II and III.  
Figure. \ref{Fig:AntennaPower} indicates that the triple-output TOBA has no blind direction and 
that its sensitivity is much more uniform than the conventional single-output TOBA. 
 
When we take $\rho_{\text{min}} = D = 1$ and assume that the sensitivity of the three signals are the same for convenience, 
then $V_1 = 1.2$ with the single signal, and $V_3 = 2.0$ with the three signals.
When we obtain three signals with the same sensitivity, the detection volume will be about 1.7 times 
larger than the single output configuration.

\section{Parameter resolution}\label{Sec:Para_resol}
In the following, we study how accurately we could measure the source location 
using a single and a network of triple-output TOBAs 
and also examine the measurement errors of the other waveform parameters.
For this purpose we consider two nominal sources, monochromatic sources and IMBH binary coalescences.  
\subsection{Fisher analysis}
Before proceeding with calculations, 
we provide a brief review of the Fisher analysis 
to evaluate parameter estimation errors for a network of $N_a$ antennas 
each of which has triple outputs (see \cite{Finn:1992wt, Cutler:1994ys, Jaranowski:2007pe} for more details). 
Each output $s_N \left(t\right)$ is assumed to be written as 
a sum of the noise $n_N \left(t\right)$ in the $N$-th output 
and the GW signal $h_N \left(t\right)$, 
$s_N \left(t\right) = h_N \left(t\right) + n_N \left(t\right)$. 
If the noise is stationary, 
the correlation between the Fourier components of the noise can be expressed by 
\begin{align} 
 \langle \tilde{n}_N \left(f\right)  \tilde{n}_M^{\ast} \left(f'\right) \rangle
 = \dfrac{1}{2}\delta\left(f-f'\right) S_n\left(f\right)_{NM}
\end{align}
where  $\langle \cdots \rangle$ denotes the ensemble average and 
$\boldsymbol{S}_n \left(f\right)$ is a one-sided power spectral 
density matrix of the antenna network. It is convenient to introduce 
a noise-weighted inner product between two $N_o$-dimensional vector 
functions $\boldsymbol{f}\left(t\right)$ and $\boldsymbol{g}\left(t\right)$, 
\begin{align}
  \left( \boldsymbol{f} \Big| \boldsymbol{g} \right)
  = 4 \text{Re} \sum_{N,M=1}^{N_o} \int_0^{\infty}  df
    \dfrac{\tilde{f}_N \left(f\right) \tilde{g}^{\ast}_M \left(f\right) }{S_n\left(f\right)_{NM}}
    \label{Eq:inner_product_def}
\end{align}
where $\text{Re}$ denotes the real part and $N_o$ is the number of the total outputs of the network. 
Using this inner product, the SNR for the GW signal $\boldsymbol{h} \left(t\right)$ can 
be written as \begin{align}
 \dfrac{S}{N} = \left( \boldsymbol{h} \Big|  \boldsymbol{h} \right)^{1/2}. \label{Eq:SNR_def}
\end{align}
We assume that the GW signal $h\left(t\right)$ is characterized by a collection of unknown parameters 
$\boldsymbol{\lambda} = \left\{ \lambda_1, \cdots, \lambda_n \right\}$. 
When the noise is Gaussian in addition to stationary, 
the statistical errors caused by the randomness of the antenna noises are estimated by 
\begin{align}
 \langle \Delta \lambda^i \Delta \lambda^j \rangle = \left(\Gamma^{-1}\right)_{ij} \label{Eq:measurement_error_def}
\end{align}
for large $S/N$. 
The matrix $\Gamma_{ij}$ is referred to as the Fisher information matrix defined by
\begin{align}
 \Gamma_{ij} 
  = \left( \dfrac{\partial \boldsymbol{h}}{\partial \lambda^i} \Big| \dfrac{\partial \boldsymbol{h}}{\partial \lambda^j} \right).  \label{Eq:Fisher_matrix_def}
\end{align}
To estimate the angular resolution of the antennas, we use the error in solid angle defined by 
\begin{align}
 \Delta \Omega 
 \equiv 2\pi \left| \sin \delta \right| 
            \sqrt{\langle \Delta \alpha^2 \rangle \langle \Delta \delta^2 \rangle - \langle \Delta \alpha \Delta \delta \rangle^2}. \label{Eq:angular_resolution_def}
\end{align}

Throughout this paper, we asssume that the noises of each output are 
uncorrelated with each other and that the form of their spectral densities are the same. 
In this case, the spectrum matrix $\boldsymbol{S}_n \left(f\right) $ 
has only diagonal elements all of which can be written with the same function $S_n\left(f\right)$. 

In practice, it is necessary to reveal and solve some technical issues in actual antennas.
For example, the loss angle of the suspension wire has to be investigated in the $\phi$ direction.
Though we assumed that the sensitivities of all three signals are the same, 
the sensitivities should be different because of the suspension thermal noise in $s_{\text{II}}$ and $s_{\text{III}}$ 
which is considered to be larger than that in $s_{\text{I}}$.

Also, Newtonian noise is one of the largest problem for low-frequency GW detection on the ground.
Newtonian noise is the fluctuation of the test masses caused by the gravity gradient of the seismic or acoustic waves. 
It would affect the sensitivity of TOBA below $f=0.1$ Hz \cite{Saulson:1984yg, Ando:2010zz} 
in the same way as for the interferometric GW detectors.
It will degrade the SNR since the Newtonian noise will appear in all the three outputs with correlations.
We have to subtract the Newtonian noise, for example, using seismic sensor arrays.

\subsection{Monochromatic sources}
\subsubsection{GW waveform}
We consider monochromatic GW with a frequency $f_0$ which can be written as 
\begin{subequations}
\begin{align}
 &h_+ \left(t\right) = A_+ \cos \left(2\pi f_0 t \right), \label{Eq:GW_monochromatic_plus}\\
 &h_\times \left(t\right) = A_\times \sin \left(2\pi f_0 t \right) \label{Eq:GW_monochromatic_cross}
\end{align}
\end{subequations}
where $A_+$ and $A_\times$ are the amplitudes of the two independent polarizations. 
When the source is a Newtonian circular binary composed of two point masses, the amplitudes are given by 
\begin{subequations}
\begin{align}
 &A_+ =  \dfrac{4G\mu \omega_s R^2}{c^4 r} \dfrac{1 +\cos ^2 \iota }{2}, \\
 &A_\times = \dfrac{4G\mu \omega_s R^2 }{c^4 r} \cos \iota
\end{align}
\end{subequations}
where $\iota$, $\mu$, $R$, $\omega_s$ and $r$ is 
the inclination, the reduced mass, the orbital radius, the orbital frequency 
and the distance to the souce, respectively (see e.g. \cite{Maggiore:2007}). 

For monochromatic GW sources, 
the relative motion of the antenna with respect to them causes the Doppler shift
which induces the time-dependence in the observed frequency. 
In order to correct the Doppler effect, 
we adopt the Solar System Barycenter (SSB) as the reference point where the time $t$ is measured.  
The Doppler correction is given by the derivative of the R\"{o}mer time delay which 
is defined as the delay between the arrival times of the antenna and the SSB; 
\begin{align}
 \Delta t = \dfrac{ \boldsymbol{n}_0 \cdot \boldsymbol{r}_d }{c} \label{Eq:doppler_correction}
\end{align}
where 
$\boldsymbol{r}_d\left(t\right)$ is the relative position vector 
pointing from the SSB to the antenna and 
$\boldsymbol{n}_0$ is the unit vector pointing from the SSB to the GW source. 
The inner product $\boldsymbol{r}_d\cdot \boldsymbol{n}_0$ is written as 
\begin{align}
   \boldsymbol{n}_0 \cdot \boldsymbol{r}_d 
 &= R_{\text{ES}} \left[  \cos \alpha \cos \delta \cos \left( \phi_o + \Omega_o t \right) + \left( \cos \varepsilon \sin \alpha \cos \delta 
    + \sin \varepsilon \sin \delta \right) \sin \left( \phi_o + \Omega_o t \right) \right] \nonumber \\
 &\hspace{1em} + R_{\text{E}} \left[\sin \lambda \sin \delta + \cos \lambda \cos \delta \cos \left( \alpha - \phi_r - \Omega_r t \right) \right]
\end{align}
where $R_{\text{ES}}$ is the distance between the SSB and the center of the Earth, 
$R_{\text{E}}$ is the Earth's radius, 
$\Omega_o$ is the angular speed of the Earth's revolution around the Sun, 
$\phi_o $ is the initial phase which defines the Earth's position at $t=0$ and 
$\varepsilon$ is the Earth's axial tilt \cite{Jaranowski:1998qm}. 
Here we neglect the Einstein time delay and the  
Shapiro time delay because these effects 
are unobservable at low-frequencies less than $1$ Hz. 

Combining Eqs.(\ref{Eq:GW_signal}), (\ref{Eq:GW_monochromatic_plus}), 
(\ref{Eq:GW_monochromatic_cross}) and (\ref{Eq:doppler_correction}), 
we find the antenna response of the $N$-th output to a monochromatic source as 
\begin{align}
 h_N \left(t\right)= A_N \left(t\right) \cos \left[ \Phi_N \left(t\right) \right] \label{Eq:GW_monochromatic_h}
\end{align}
where the suffix ``$N$" stands for the $N$-th output and 
the function $A_N \left(t\right), Q_N\left(t\right)$ and  $\Phi_N \left(t\right)$ are given as follows. 
\begin{subequations} 
\begin{align}
 &A_N\left(t\right) 
   =  \mathcal{A} Q_N \left(t\right), \label{Eq:GW_monochromatic_A}\\
 &Q_N\left(t\right) 
   =  \left[ \left(\dfrac{1+\cos^2 \iota}{2}\right)^2 F_N^{+}{}^2\left(t\right) +  \cos^2\iota F_N^{\times}{}^2 \left(t\right) \right]^{1/2}, \label{Eq:GW_monochromatic_Q}\\
 &\Phi_N \left(t\right) 
   = 2\pi f_0 t + \varphi_0 + \varphi_{p,N} \left(t\right) + \varphi_D \left(t\right),  \\
 &\varphi_{p,N} \left(t\right) 
   =  \arctan\left[ - \dfrac{2\cos\iota}{1+\cos^2\iota}\dfrac{ F^\times_N \left(t\right) }{F_N^+\left(t\right) } \right], \label{Eq:GW_monochromatic_phi_p}\\
 &\varphi_D \left(t\right) 
   =  2\pi f_0  \dfrac{\boldsymbol{n}_0 \cdot \boldsymbol{r}_d \left(t\right) }{c} \label{Eq:GW_monochromatic_phi_D}
\end{align}
\end{subequations}
where $\mathcal{A}$ is the overall amplitude and $\varphi_0$ is a constant reference phase. 
The phase shift $\varphi_{D} \left(t\right)$ which appears in the GW phase $\Phi\left(t\right)$ 
is called the Doppler phase because it comes from the Doppler correction  (\ref{Eq:doppler_correction}). 
The phase shift $\varphi_{p,N} \left(t\right)$ is the  polarization phase 
which depends on the angular pattern functions $F_+\left(t\right)$ and $F_\times\left(t\right)$. 
\subsubsection{Parameter resolution}
The GW signal from the monochromatic source 
Eqs.(\ref{Eq:GW_monochromatic_h})-(\ref{Eq:GW_monochromatic_phi_D}) 
is characterized by the seven waveform parameters, 
the overall amplitude, $\mathcal{A}$,  
the GW frequency, $f_0$,  
the reference phase, $\varphi_0$,  
the sky position of the source, $\alpha$ and $\delta$,  
the polarization angle, $\psi$,  
the inclination, $\iota$. 
Before proceeding we simplify the task of integration in the Fisher matrix (\ref{Eq:Fisher_matrix_def}).  
The time scale of change in $\Phi_N\left(t\right)$ is of the order of $f_0$. 
Meanwhile, the temporal change in  $A_N\left(t\right) $ is mainly caused by the Earth's rotation.  
So the amplitude $A_N\left(t\right)$ varies slowly with time 
in comparison with the phase $\Phi_N\left(t\right)$ 
and the inequality $f_0 \gg \left|dA_N /dt\right|/A_N$ holds. This leads to 
\begin{subequations}
\begin{align}
 &\left(\dfrac{S}{N}\right)^2 
   \cong \dfrac{1}{S_n\left(f_0 \right)} \sum_{N} \int_0^{T_{\text{obs}}} dt \ A_N\left(t\right)^2, \label{Eq:SN_monochromatic}\\
 &\Gamma_{ij} 
   \cong  \dfrac{1}{S_n\left(f_0 \right)} \sum_{N} \int_0^{T_{\text{obs}}} dt \ \left[\partial_i A_N\left(t\right) 
   \partial_j A_N\left(t\right) + A_N\left(t\right)^2 \partial_i \Phi_N\left(t\right)  \partial_j \Phi_N\left(t\right) \right] \label{Eq:Fisher_monochromatic}
\end{align} 
\end{subequations}
from Eqs.(\ref{Eq:SNR_def}) and (\ref{Eq:Fisher_matrix_def}) 
where the rapidly oscillating terms in the integrand such as $\cos \Phi_N \left(t\right)$ and $\sin \Phi_N \left(t\right)$ 
are discarded and  $T_{\text{obs}}$ denotes the observation time \cite{Cutler:1997ta, Takahashi:2002ky}. 
Observing that $S_n \left(f_0\right)$ in Eq.(\ref{Eq:Fisher_monochromatic}) 
can be eliminated by Eq.(\ref{Eq:SN_monochromatic}), 
the measurement errors can be expressed in terms of $S/N$ which is normalized by $S/N=10$ in this section.  
Note that the measurement errors are independent of the form of $S_n \left(f\right)$. 
Making the substitution of  Eqs.(\ref{Eq:GW_monochromatic_A})-(\ref{Eq:GW_monochromatic_phi_D}) 
into Eq.(\ref{Eq:Fisher_monochromatic}) 
produces the values of the Fisher matrix elements $\Gamma_{ij}$. 
Using Eq.(\ref{Eq:measurement_error_def}), we obtain the measurement errors. 
 
\begin{figure}[htbp] 
\centering 
\includegraphics[width=8cm,clip]{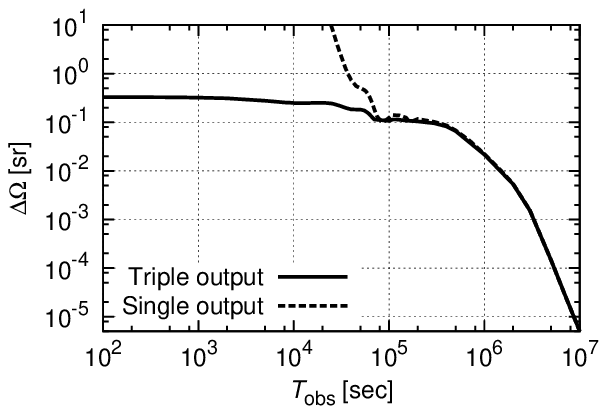}
\caption{ 
Angular resolution $\Delta \Omega$ vs the observation time $T_{\text{obs}}$ 
for monochromatic sources with $f_0 = 1 \ \text{Hz}$ using a  single antenna. 
We take the initial position and orientation of the antenna and the source position to be 
$\phi_r = \phi_o = 0, \alpha= \delta = \iota = \psi = 1.0$ radian. 
GW signals are normalized by $S/N=10$.  
The solid line and the dashed line correspond to a three-output TOBA and a one-output TOBA respectively. }
\label{Fig:error_f=1.0_deltaOmega_single}
\centering 
\includegraphics[width=8cm,clip]{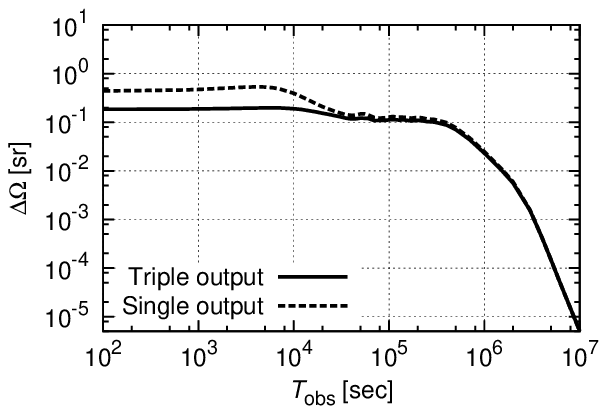}
\caption{
Angular resolution $\Delta \Omega$ vs the observation time $T_{\text{obs}}$ 
for monochromatic sources with $f_0 = 1 \ \text{Hz}$ using three antennas. 
We take the initial position and orientation of the antenna and the source position to be 
$\phi_o = 0, \alpha= \delta = \iota = \psi = 1.0$ radian. 
GW signals are normalized by $S/N=10$.  
The solid line and the dashed line correspond to a three-output TOBA and a one-output TOBA respectively. }
\label{Fig:error_f=1.0_deltaOmega_multi}
\end{figure}

Figure. \ref{Fig:error_f=1.0_deltaOmega_single} shows the angular resolution 
for a monochromatic source with a frequency $f_0=1 \ \text{Hz}$ 
using a single antenna located at the TAMA300's site in Japan. 
This result is normalized by $S/N=10$ with parameters $\phi_r = \phi_o = 0, \alpha= \delta = \iota = \psi = 1.0$ radian. 
The two curves, the dashed line and the solid line  
correspond to the antennas which have one output and three outputs, respectively. 
As can be seen in Fig. \ref{Fig:error_f=1.0_deltaOmega_single}, 
there is no difference between two curves 
for $T_{\text{obs}} > 1 $ day 
because the angular resolution is mainly determined by the Doppler phase $\varphi_{D}$. 
When $T_{\text{obs}} > 1 $ day, the error $\Delta \Omega$ drops with the observation time for $T_{\text{obs}} > 10^5$ seconds 
and approaches to the diffraction-limited accuracy \cite{Schutz:2001re}.  
This feature can be traced to the Doppler shift caused by the revolution around the Sun.  
The three-output antenna can be treated as the same way 
as the one-output antenna for such a long-duration signals with the fixed $S/N$. 
The difference between two curves appears 
in the short-duration signals such as $T_{\text{obs}} < 1 $ day. 
The errors of the sky position for the one-output antenna decrease with time 
for $T_{\text{obs}} \gtrsim 10^4$ seconds 
due to the Doppler phase caused by the Earth's rotation.  
However, the one-output antenna cannot determine the source position 
for $T_{\text{obs}} \lesssim 10^4$ seconds at all. 
On the other hand, the angular resolution for the three-output antenna
is of the order of $0.1$ steradians  
and remains approximately constant 
regardless of the observation time less than 1 day. 
This feature can be traced to the polarization phase $\varphi_{p,N}$. 
For the three-output antenna, the two independent polarization modes, plus mode and cross mode 
can be discriminated because the three independent signals can be obtained even with a single antenna. 
The degeneracy of the two polarization modes 
is resolved by the three-output antenna unlike the one-output antenna. 
Thus, the angular resolution of the three-output antenna for short-duration signals 
is much better than that of the one-output antenna thanks to the information on the polarization. 
Note that a mirror-image ambiguity remains in the location on the sky with a single antenna observation. 
Two or more antennas are required for resolving this ambiguity. 

Figure. \ref{Fig:error_f=1.0_deltaOmega_multi} shows the angular resolution 
for the monochromatic source with the frequency $f_0=1.0 \ \text{Hz}$ using a network of three antennas. 
This result is normalized by $S/N=10$. 
Each antenna is located at the site of TAMA300 (Japan), LIGO-Hanford (US) and Virgo (Italy). 
The two curves, the dashed line and the solid line 
correspond to the antennas which have one output and three outputs, respectively.
As can be seen in Fig. \ref{Fig:error_f=1.0_deltaOmega_multi}, 
the two curves coincide with each other for long-duration signals. 
This behavior is the same as the case for the observations with a single antenna for $T_{\text{obs}} > 1$ day. 
On the other hand, the accuracy of the source location for a network of  one-output antennas  
is constant and of the order of 0.1 steradians below $T_{\text{obs}}\sim 10^4 $ seconds 
because it is mainly determind by the polarization phase $\varphi_{p,N}$.
It should be noted that the angular resolution of a network of ground-based laser interferometers 
which are sensitive to an audio frequency signal is determined by the error of the delays of the arrival time \cite{Guersel:1989th}. 
It is roughly estimated by the geometrical formula derived by Wen and Chen \cite{Wen:2010cr}
\begin{align}
 \Delta \Omega
 &= 0.25
    \left( \dfrac{10 \text{Hz}}{f_0} \right)^2 \left( \dfrac{3 \times 10^{13} \text{m}^2}{A} \right)\left( \dfrac{0.5}{\cos \chi} \right)
    \left( \dfrac{10}{S/N} \right)^2 
   \left( \dfrac{\left(S/N\right)_1 
    \left(S/N\right)_2  \left(S/N\right)_3}{3\sqrt{3}\left(S/N\right)^3 } \right)^{-1}  \label{Eq:WenChen_formula}
\end{align}
where $f_0$ is the GW frequency, $A$ is the area formed by the network of the three antennas, 
$S/N$ is the total SNR, 
$\left(S/N\right)_i$ is the SNR achievable with the $i$-th antenna alone, 
and $\chi$ is the angle between the normal to the plane defined by the antennas and the direction of  GW propagation. 
The value $3\times 10^{13} \text{m}^2$ corresponds to the area formed by the TAMA-LIGO-Virgo network. 
This formula propotional to $1/f_0^2$ does not work when applied to a low frequency source because 
the directional derivatives of the pattern functions 
such as $\partial F_+ / \partial \alpha$ are neglected in the derivation of Eq. (\ref{Eq:WenChen_formula}) \cite{Wen:2010cr}.  
When the source location is identified using the network of the TOBAs which are sensitive to a sub-audio frequency signal, 
the polarization phases $\varphi_{p,N}$ but not the Doppler phase $\varphi_{D}$ play a key role in its angular resolution. 
The difference between two curves in Fig. \ref{Fig:error_f=1.0_deltaOmega_multi} 
comes from the number of the independent outputs in each antenna. 

Table \ref{Fig:parameter_accuracy_monochromatic_source_dif} lists 
the angular resolution for the various frequencies and angular parameters. 
The results in Table \ref{Fig:parameter_accuracy_monochromatic_source_dif} 
indicates the angular resolution for short-duration signals is independent of the GW frequency $f_0$ 
because it is determined by $\varphi_{p,N}\left(t\right)$ which does not depend on $f_0$. 

\begin{figure}[htbp] 
\centering 
\includegraphics[width=15cm,clip]{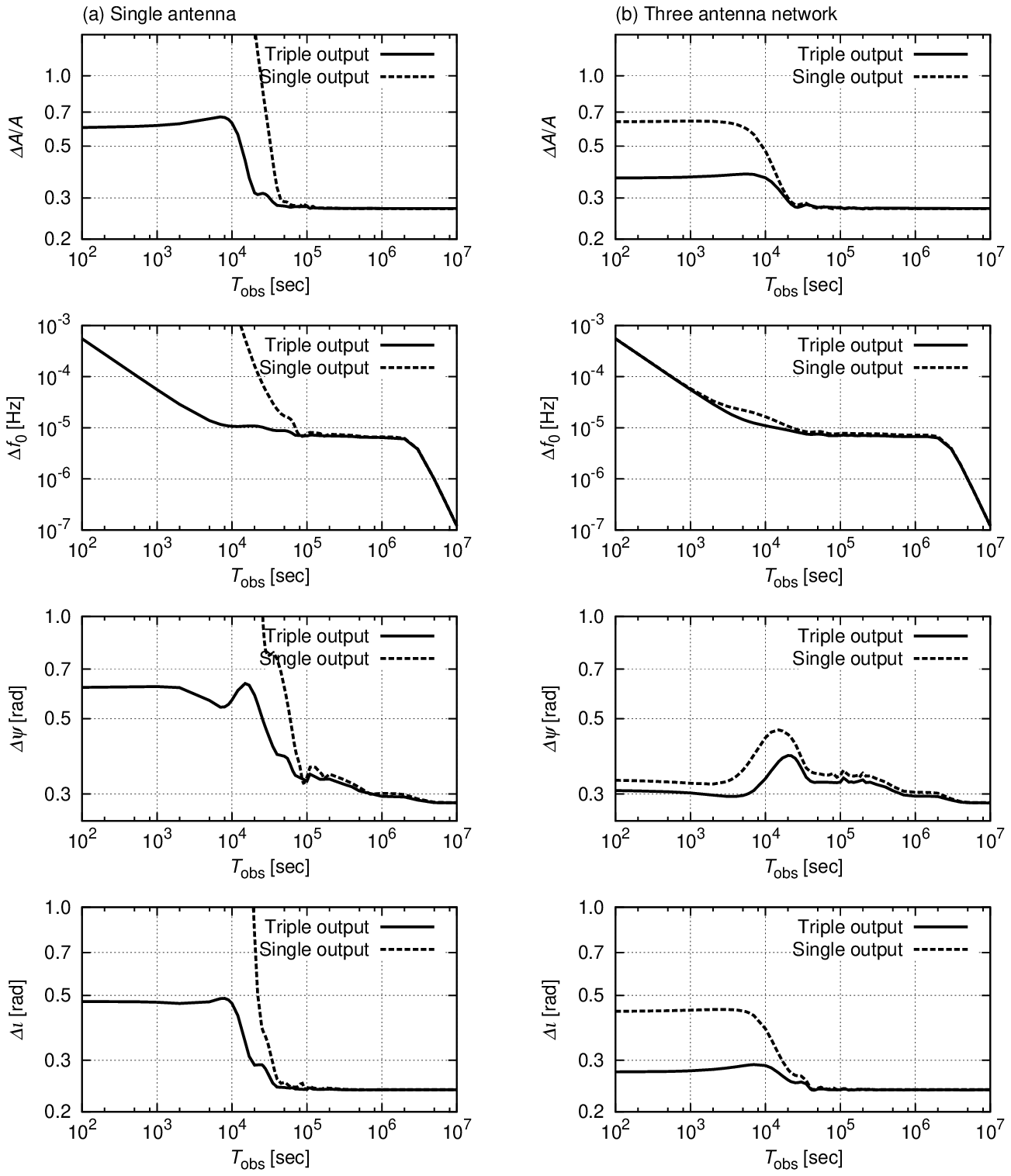}
\caption{ 
Parameter accuracies for the monochromatic source with frequency $f_0 = 1$ Hz 
are plotted vs observation time $T_{\text{obs}}$. 
The left and right column correspond to a single antenna case 
and a three antenna network case, respectively. 
All results are normalized by $S/N=10$. 
The initial position and orientation of the antenna 
at the TAMA300's site and the angular parameters 
are assume to be $\phi_r = \phi_o = 0, \alpha= \delta = \iota = \psi = 1.0$. }
\label{Fig:parameter_accuracy_monochromatic_source}
\end{figure}

The measurement errors of the other waveform parameters 
$\Delta \mathcal{A}, \Delta f_0, \Delta \psi$ and $\Delta \iota$ 
for the monochromatic wave with $f_0 = 1$ Hz 
are described in Fig. \ref{Fig:parameter_accuracy_monochromatic_source}.   
These results are also normalized by $S/N=10$. 
The behavior of the curves in Fig. \ref{Fig:parameter_accuracy_monochromatic_source} 
can essentially be explained in the same way as we discussed above. 
For short-duration signals with the observation time 
less than $10^4$ seconds, the errors are determined by 
the polarization phase $\varphi_{p,N}$ for both 
a single and a network of triple-output antennas. 
When $T_{\text{obs}} > 10^4$ seconds 
oscillating parts appear in Fig. \ref{Fig:parameter_accuracy_monochromatic_source} 
due to the phase-modulation induced by the Earth's spin.  
The triple outputs improve the parameter accuracies of the short-duration signals drastically 
for the single antenna case and improve them by several factors for the three antenna network case.   
\begin{table}[htbp]
\begin{center}
\begin{tabular}{ccccccc} \hline \hline
  $f_0 \ \left[\text{Hz}\right]$ & $\alpha \ \left[ \text{rad} \right]$ & $\delta \ \left[ \text{rad} \right]$ 
  & $\iota \ \left[ \text{rad} \right]$ & $\psi \ \left[ \text{rad} \right]$ & $\left(S/N \right)_I$ & $\Delta \Omega \ \left[\text{sr}\right]$ \\ \hline
  $1.0$  & $1.0$ & $1.0$   & $0.45$ & $1.0$ & $8.90$ & $0.332$ \\
             & $0.5$ & $-1.0$  & $0.45$ & $0.0$ & $6.82$ & $0.184$ \\
             & $1.0$ & $1.0$   & $1.0$  & $1.0$ & $8.73$ & $0.292$ \\
             & $0.5$ & $-1.0$  & $1.0$  & $0.0$ & $7.07$ & $0.187$ \\
             & $1.0$ & $1.0$   & $1.5$ & $1.0$ & $4.82$ & $0.206$ \\
             & $0.5$ & $-1.0$  & $1.5$ & $0.0$ & $8.09$ & $0.365$ \\ \hline
  $0.1$  & $1.0$ & $1.0$   & $0.45$ & $1.0$ & $8.90$ & $0.332$ \\
             & $0.5$ & $-1.0$  & $0.45$ & $0.0$ & $6.82$ & $0.184$ \\
             & $1.0$ & $1.0$   & $1.0$  & $1.0$ & $8.73$ & $0.292$ \\
             & $0.5$ & $-1.0$  & $1.0$  & $0.0$ & $7.07$ & $0.187$ \\
             & $1.0$ & $1.0$   & $1.5$ & $1.0$ & $4.82$ & $0.206$ \\
             & $0.5$ & $-1.0$  & $1.5$ & $0.0$ & $8.09$ & $0.365$ \\ \hline
  $0.01$& $1.0$ & $1.0$   & $0.45$ & $1.0$ & $8.90$ & $0.332$ \\
             & $0.5$ & $-1.0$  & $0.45$ & $0.0$ & $6.82$ & $0.184$ \\
             & $1.0$ & $1.0$   & $1.0$  & $1.0$ & $8.73$ & $0.292$ \\
             & $0.5$ & $-1.0$  & $1.0$  & $0.0$ & $7.07$ & $0.187$ \\
             & $1.0$ & $1.0$   & $1.5$ & $1.0$ & $4.82$ & $0.206$ \\
             & $0.5$ & $-1.0$  & $1.5$ & $0.0$ & $8.09$ & $0.365$ \\ \hline\hline
\end{tabular}
\end{center}
\caption{Angular resolution for monochromatic sources with frequency $f_0$ 
using the single three-output TOBA for 1-hour observation.  
The result is normalized by $S/N=10$. $\left(S/N\right)_I$ denotes 
the SNR of the output I.  
We take the value of the initial position and orientation of the antenna to be $\phi_r = \phi_o = 0$ radian. }
\label{Fig:parameter_accuracy_monochromatic_source_dif}
\end{table}
\subsection{IMBH-IMBH binary coalescence}
\subsubsection{GW waveform}
In this section, 
we consider a coalescing binary system composed of two point masses with $m_1$ and $m_2$ as a GW source. 
Unlike the monochromatic sources we discussed in the previous section, 
the orbital radius shrinks with time and 
the orbital frequency increases accordingly because of the GW radiation. 
In this case, the parameter accuracies depend on the form of the noise spectral density of the antenna $S_n\left(f\right)$. 
As a signal model from the binary coalescing, 
we adopt the restricted post-Newtonian waveform with the 1.5 PN phase 
in which the amplitude is retained up to the Newtonian order: 
\begin{subequations} 
\begin{align}
 &\tilde{h}_N\left(f\right) 
   = \mathcal{A} Q_N\left(t_{\ast} \right) f^{-7/6} 
       e^{-i\left( \varphi_{p,N}\left(t_{\ast} \right) + \varphi_D\left(t_{\ast} \right) \right)}  
       e^{i\Psi\left(f\right)}, \label{Eq:CBC_waveform_Fourier_h}\\
 &\mathcal{A} 
   \equiv \sqrt{\dfrac{5}{24}} \dfrac{1}{\pi^{2/3}} \dfrac{c}{r} \left(\dfrac{GM_c}{c^3}\right)^{5/6}, \\
 &\Psi\left(f\right) 
   = 2\pi ft_{c} - \dfrac{\pi}{4} -\phi_c- \tilde{\Phi}\left(f\right), \\
 &\tilde{\Phi}\left(f\right) 
   = -\dfrac{3}{4}\left(\dfrac{GM_c}{c^3}8\pi f \right)^{-5/3} 
      \left[ 1 + \dfrac{20}{9} \left( \dfrac{743}{336} + \dfrac{11}{4}\eta \right)x - 16\pi x^{3/2} \right], \\
 &t_{\ast}\left(f\right)
   = t_c - 5 \left(\dfrac{GM_c}{c^3}\right)^{-5/3} \left(8 \pi f\right)^{-8/3}
      \left[ 1 + \dfrac{4}{3} \left( \dfrac{743}{336} + \dfrac{11}{4}\eta \right)x - \dfrac{32}{5}\pi x^{3/2}  \right] \label{Eq:CBC_waveform_Fourier_t}
\end{align}
\end{subequations}
where $r$, $t_c$ and $\phi_c$ are 
the distance to the source, the time and the phase at the coalescence, respectively (see e.g. \cite{Maggiore:2007}).  
The masses $M_c \equiv \left(m_1 m_2 \right)^{3/5} \left(m_1 + m_2 \right)^{-1/5}$ 
and  $\eta \equiv m_1 m_2 /\left(m_1 + m_2 \right)^2$ 
are called a chirp mass and a symmetric mass ratio respectively and 
a PN variable $x\equiv \left[ \pi G \left(m_1 + m_2\right) f/c^3  \right]^{2/3}$ was introduced.
We cut-off the GW signal at $f_{\text{max}} \equiv c^3 / 6\sqrt{6}\pi G\left(m_1 + m_2\right)$ 
beyond which the circular orbit is not stable any longer and the two point masses plunge toward each other. 
So we set $\tilde{h} \left(f\right) = 0$ for $f > f_{\text{max}}$ in Eq.(\ref{Eq:CBC_waveform_Fourier_h}). 

\subsubsection{Parameter resolution}
We analyze the angular resolution for the coalescecing IMBH binary signal 
using Eqs. (\ref{Eq:CBC_waveform_Fourier_h})-(\ref{Eq:CBC_waveform_Fourier_t}). 
The GW signal is described by eight waveform parameters, 
the overall amplitude, $\mathcal{A}$, 
the coalescence time, $t_c$, 
the coalescence phase, $\phi_c$,  
the sky position of the source, $\alpha$ and $\delta$,  
the polarization angle, $\psi$, 
the inclination, $\iota$, and 
the chirp mass, $M_c$. 
Substitution of Eq. (\ref{Eq:CBC_waveform_Fourier_h}) 
into Eq.(\ref{Eq:Fisher_matrix_def}) yields 
the values of Fisher matrix elements. 
From Eq.(\ref{Eq:measurement_error_def}) we get the accuracies of the waveform parameters. 
  
We present a representative examples of our result 
in Table \ref{Table:CBC_deltaOmega_single} and \ref{Table:CBC_deltaOmega_multi} 
where source parameters are chosen  randomly.
We set the coalescence binary at $100$ Mpc and the initial position of the antenna located 
at the TAMA300's site to be $\phi_r = \phi_o = 0$. 
Here, we do not assume the fixed SNR  and neglect the effect of the expansion of the Universe and 
the higher order terms in the Post-Newtonian expansion.   
However, the angular resolution $\Delta \Omega$ we calculate in this paper  
is expected to be accurate because the angular resolution is accumulated 
at the inspiral phase long before the final plunge. 

The $S/N$ and the angular resolution $\Delta \Omega$ 
for the single antenna case (the three antenna network case) 
are listed in Table \ref{Table:CBC_deltaOmega_single} (Table \ref{Table:CBC_deltaOmega_multi}).  
The result of the single antenna case listed in Table \ref{Table:CBC_deltaOmega_single}
shows that the three-output antenna improves the $S/N$ 
by a factor of about $1.2$ in comparison with the one-output antenna. 
The same can be said of the three anntenna network listed in Table \ref{Table:CBC_deltaOmega_multi}. 
This factor can be simply explained by 
$\sqrt{1^2 + \left(1/2\right)^2 + \left(1/2\right)^2}\cong 1.22 $ 
from Eqs.(\ref{Eq:def_hI})-(\ref{Eq:def_hIII}). 
As can be seen in Table \ref{Table:CBC_deltaOmega_single}, 
the angular resolution of the single three-output antenna $\Delta \Omega$ ranges from $10^{-2}$ to 1 steradians. 
It is roughly an order of magnitude better than that of the one-output antenna. 
This is because the angular resolution for short-duration signals such as signals from coalescing binaries  
is mainly determined by the polarization phase $\varphi_{p,N}\left(t\right)$ but not 
the Doppler phase $\varphi_D \left(t\right)$ as we discussed in the previous section 
where the monochromatic source was investigated. 
Since the two polarization modes (plus and cross modes) are degenerate in 
the single antenna which has the only one output, 
it is hard to locate the GW sources for short-duration signals. 
However, the single three-output antenna can identify the source location to some degree 
because the degeneracy of the two polarization modes is resolved. 
Table \ref{Table:CBC_deltaOmega_multi} shows that 
the angular resolution of the network of three-output antennas 
ranges roughly from 0.1 to $10^{-3}$ steradians. 
It is a factor of about 2 better than that of the one-output antenna network. 
This improvement is accomplished mainly by 
increasing $S/N$ which comes from the triple outputs. 
The measurement errors of the other waveform parameters such as $\Delta \mathcal{A}/\mathcal{A}$ 
are also improved by several factors due to the additional outputs. 
 \begin{table}[htbp] 
\begin{center}
\begin{tabular}{cccccc|cccccccccc} \hline\hline
  $m_1$                          & $m_2$        & $\alpha$                     & $\delta $       & $\iota$                           & $\psi$ &
  $\left(S/N \right)_{\text{I}}$ & $S/N$        & $\Delta \Omega_{\text{I}}$   & $\Delta \Omega$ & $\Delta\ln\mathcal{A}_{\text{I}}$ & $\Delta\ln\mathcal{A}$
  & $\Delta\iota_I$              & $\Delta\iota$&$\Delta\psi_I$                 & $\Delta\psi$  \\
  $\left[ M_{\odot} \right]$     & $ \left[ M_{\odot} \right]$ & $\left[ \text{rad}\right]$ & $\left[ \text{rad}\right]$ & $\left[ \text{rad}\right]$ & $\left[ \text{rad}\right]$ &
  &    & $ \left[\text{sr}\right]$ & $ \left[\text{sr}\right]$ & &  & $\left[ \text{rad}\right]$ & $\left[ \text{rad}\right]$ & $\left[ \text{rad}\right]$ & $\left[ \text{rad}\right]$ \\ \hline
  $10^4$ & $10^4$ 
            & $1.0$   & $1.0$  & $1.0$ & $1.0$ & $29.9$ &  $35.3$ & $4.02$  & $0.0253$& 1.76 & 0.165 & 0.976& 0.132 & 1.67 & 0.172    \\
         &  & $2.0$   & $-1.0$ & $0.5$ & $3.0$ & $46.1$ &  $54.3$ & $2.70$  & $0.0207$& 3.75 & 0.631 & 7.44 & 1.25  & 0.833& 2.60     \\
         &  & $1.0$   & $-1.5$ & $0.5$ & $4.0$ & $40.1$ &  $50.6$ & $23.2$  & $0.258$ & 3.97 & 0.697 & 8.05 & 1.38  & 13.6 & 2.38    \\
         &  & $-3.0$  & $0.5$  & $1.0$ & $0.0$ & $25.0$ &  $32.2$ & $0.765$ & $0.0265$& 2.17 & 0.246 & 1.92 & 0.178 & 2.16 & 0.198   \\
         &  & $3.0$   & $1.5$  & $1.0$ & $2.0$ & $23.4$ &  $31.9$ &  $47.2$ & $0.348$ & 1.40 & 0.166 & 1.00 & 0.143 & 12.3 & 1.11    \\
         &  & $-1.0$  & $1.0$  & $0.5$ & $6.0$ & $48.6$ &  $55.8$ &  $2.90$ & $0.0188$& 3.94 & 0.652 & 7.60 & 1.23  & 13.4 & 2.52   \\
         &  & $-2.0$  & $-0.5$ & $1.0$ & $5.0$ & $26.8$ & $34.4$ &  $1.47$  & $0.0757$& 2.13 & 0.261 & 1.39 & 0.252 & 1.39 & 0.409  \\
         &  & \multicolumn{2}{c}{All-sky average} & $1.0$ & $1.0$ & $25.9$ & $33.2$ &  $6.82$  & $0.0674$& 1.46 & 0.216 & 1.30 & 0.181 & 2.61 & 0.285   \\ \hline
  $10^4$  & $10^5$
            & $1.0$   & $1.0$  & $1.0$ & $1.0$ & $10.3$ &  $12.3$ & $4.18$  & $0.180$ & 1.80 & 0.414 & 1.04 & 0.339 & 1.67 & 0.448   \\
         &  & $2.0$   & $-1.0$ & $0.5$ & $3.0$ & $16.4$ &  $19.2$ & $2.93$  & $0.127$ & 3.79 & 1.54  & 7.49 & 3.03  & 14.4 & 6.36  \\
         &  & $1.0$   & $-1.5$ & $0.5$ & $4.0$ & $14.1$ &  $17.8$ & $24.8$  & $1.58$  & 4.08 & 1.70  & 8.24 & 3.38  & 14.2 & 5.69   \\
         &  & $-3.0$  & $0.5$  & $1.0$ & $0.0$ & $8.77$ &  $11.4$ & $0.846$ & $0.170$ & 2.30 & 0.566 & 2.04 & 0.424 & 2.18 & 0.518   \\
         &  & $3.0$   & $1.5$  & $1.0$ & $2.0$ & $8.21$ &  $11.2$ & $49.9$  & $2.35$  & 1.47 & 0.444 & 1.08 & 0.370 & 12.4 & 2.78   \\
         &  & $-1.0$  & $1.0$  & $0.5$ & $6.0$ & $17.3$ &  $19.7$ & $3.12$  & $0.120$ & 4.10 & 1.61  & 7.92 & 3.05  & 13.8 & 6.28   \\
         &  & $-2.0$  & $-0.5$ & $1.0$ & $5.0$ & $9.11$ &  $11.9$ & $1.58$  & $0.311$ & 2.32 & 0.551 & 1.52 & 0.519 & 1.49 & 0.790   \\ 
         &  & \multicolumn{2}{c}{All-sky average} & $1.0$ & $1.0$ & $9.12$ & $11.7$ &  $7.26$  & $0.383$& 1.54 & 0.490 & 1.37 & 0.416 & 2.73 & 0.673  \\ \hline
  $10^5$ & $10^5$ 
            & $1.0$   & $1.0$  & $1.0$ & $1.0$ & $13.4$ &  $16.0$ & $4.11$  & $0.113$ & 1.75 & 0.336 & 1.00 & 0.273 & 0.448 & 0.358   \\
         &  & $2.0$   & $-1.0$ & $0.5$ & $3.0$ & $21.1$ &  $24.8$ & $2.82$  & $0.0841$& 3.72 & 1.26  & 7.35 & 2.49  & 6.36  & 5.21  \\
         &  & $1.0$   & $-1.5$ & $0.5$ & $4.0$ & $18.2$ &  $23.0$ & $24.2$  & $1.05$  & 3.97 & 1.40  & 8.02 & 2.77  & 5.69  & 4.70    \\
         &  & $-3.0$  & $0.5$  & $1.0$ & $0.0$ & $11.3$ &  $14.7$ & $0.804$ & $0.112$ & 2.28 & 0.475 & 2.02 & 0.352 & 0.518 & 0.415   \\
         &  & $3.0$   & $1.5$  & $1.0$ & $2.0$ & $10.6$ &  $14.4$ &  $48.1$ & $1.51$  & 1.39 & 0.352 & 1.03 & 0.297 & 2.78  & 2.26   \\
         &  & $-1.0$  & $1.0$  & $0.5$ & $6.0$ & $22.2$ &  $25.4$ &  $2.99$ & $0.0785$& 3.91 & 1.32  & 7.57 & 2.48  & 6.28  & 5.11   \\
         &  & $-2.0$  & $-0.5$ & $1.0$ & $5.0$ & $11.9$ &  $15.5$ &  $1.55$ & $0.238$ & 2.19 & 0.475 & 1.43 & 0.451 & 0.790 & 0.703   \\
         &  & \multicolumn{2}{c}{All-sky average} & $1.0$ & $1.0$ & $11.8$ & $15.1$ &  $7.05$  & $0.258$&  1.50 & 0.406 &  1.34 & 0.344 & 2.68 & 0.553   \\  \hline\hline
\end{tabular}
\end{center}
\caption{
Angular resolution for IMBH-IMBH binary coalescence signals from a distance of 100 Mpc 
using the single antenna on the TAMA300's site.  
$\left(S/N\right)_{\text{I}}$ denotes 
the SNR of the output I defined by (\ref{Eq:def_hI}).  
$S/N$ denotes the total SNR of the single antenna.  
}
\label{Table:CBC_deltaOmega_single}
\begin{center}
\begin{tabular}{cccccc|cccccccccc} \hline\hline
  $m_1$                          & $m_2$        & $\alpha$                     & $\delta $       & $\iota$                           & $\psi$ &
  $\left(S/N \right)_{1}$ & $S/N$        & $\Delta \Omega_{1}$   & $\Delta \Omega$ & $\Delta\ln\mathcal{A}_{1}$ & $\Delta\ln\mathcal{A}$
  & $\Delta\iota_1$              & $\Delta\iota$&$\Delta\psi_1$                & $\Delta\psi$  \\
  $\left[ M_{\odot} \right]$     & $ \left[ M_{\odot} \right]$ & $\left[ \text{rad}\right]$ & $\left[ \text{rad}\right]$ & $\left[ \text{rad}\right]$ & $\left[ \text{rad}\right]$ &
  &    & $ \left[\text{sr}\right]$ & $ \left[\text{sr}\right]$ & &  & $\left[ \text{rad}\right]$ & $\left[ \text{rad}\right]$ & $\left[ \text{rad}\right]$ & $\left[ \text{rad}\right]$ \\ \hline
  $10^4$& $10^4$ 
             & $1.0$   & $1.0$  & $1.0$ & $1.0$ & $48.5$ & $59.2$   & $0.0165$ & $0.00514$ & 0.122 & 0.0613 & 0.0852 & 0.0460 & 0.696  & 0.0524   \\
          &  & $2.0$   & $-1.0$ & $0.5$ & $3.0$ & $76.7$ & $89.9$   & $0.00404$& $0.00187$ & 0.399 & 0.270  & 0.777  & 0.524  & 1.64   & 1.10     \\
          &  & $1.0$   & $-1.5$ & $0.5$ & $4.0$ & $78.5$ &  $93.0$  & $0.0373$ & $0.0212$  & 0.305 & 0.204  & 0.578  & 0.401  & 1.24   & 0.819    \\
          &  & $-3.0$  & $0.5$  & $1.0$ & $0.0$ & $47.9$ &  $59.9$  & $0.00232$& $0.00181$ & 0.0639& 0.0514 & 0.0579 & 0.0467 & 0.0729 & 0.0610   \\
          &  & $3.0$   & $1.5$  & $1.0$ & $2.0$ & $54.3$ &  $64.1$  & $0.0831$ & $0.0440$  & 0.0716& 0.0500 & 0.0767 & 0.0511 & 0.462  & 0.336    \\
          &  & $-1.0$  & $1.0$  & $0.5$ & $6.0$ & $77.3$ &  $90.5$  & $0.00536$& $0.00209$ & 0.426 & 0.267  & 0.797  & 0.502  & 1.65   & 1.04     \\
          &  & $-2.0$  & $-0.5$ & $1.0$ & $5.0$ & $44.3$ &  $56.9$  & $0.00456$& $0.00210$ & 0.114 & 0.0653 & 0.0891 & 0.0520 & 0.129  & 0.0724   \\ 
         &  & \multicolumn{2}{c}{All-sky average} & $1.0$ & $1.0$ & $49.1$ & $59.3$ &  $0.0248$  & $0.00744$& 0.131 & 0.0645 & 0.114 & 0.0555 & 0.177 & 0.0931   \\ \hline
  $10^4$& $10^5$
             & $1.0$   & $1.0$  & $1.0$ & $1.0$ & $17.0$ & $20.8$   & $0.106$  & $0.0393$  & 0.311 & 0.170  & 0.219 & 0.128  & 0.201 & 0.153  \\
          &  & $2.0$   & $-1.0$ & $0.5$ & $3.0$ & $27.0$ & $31.6$   & $0.0302$ & $0.0146$  & 1.05  & 0.734  & 2.04  & 1.42   & 4.21  & 2.93   \\
          &  & $1.0$   & $-1.5$ & $0.5$ & $4.0$ & $27.5$ &  $32.6$  & $0.276$  & $0.166$   & 0.824 & 0.584  & 1.56  & 1.11   & 3.24  & 2.21   \\
          &  & $-3.0$  & $0.5$  & $1.0$ & $0.0$ & $16.8$ &  $21.0$  & $0.0203$ & $0.0155$  & 0.182 & 0.147  & 0.165 & 0.134  & 0.211 & 0.177  \\
          &  & $3.0$   & $1.5$  & $1.0$ & $2.0$ & $19.1$ &  $22.5$  & $0.601$  & $0.343$   & 0.192 & 0.139  & 0.208 & 0.143  & 1.24  & 0.935  \\
          &  & $-1.0$  & $1.0$  & $0.5$ & $6.0$ & $27.2$ &  $31.8$  & $0.0386$ & $0.0163$  & 1.08  & 0.711  & 2.03  & 1.35   & 4.26  & 2.81   \\
          &  & $-2.0$  & $-0.5$ & $1.0$ & $5.0$ & $15.6$ &  $20.0$  & $0.0366$ & $0.0172$  & 0.307 & 0.181  & 0.242 & 0.145  & 0.337 & 0.201  \\
         &  & \multicolumn{2}{c}{All-sky average} & $1.0$ & $1.0$ & $17.2$ & $20.8$ &  $0.133$  & $0.0571$& 0.305 & 0.178 & 0.267 & 0.153 & 0.425 & 0.258  \\  \hline
  $10^5$& $10^5$  
             & $1.0$   & $1.0$  & $1.0$ & $1.0$ & $22.0$ & $26.9$   & $0.0691$ & $0.0241$  & 0.250 & 0.133  & 0.76  & 0.100  & 0.155 & 0.117  \\
          &  & $2.0$   & $-1.0$ & $0.5$ & $3.0$ & $34.9$ & $40.8$   & $0.0187$ & $0.00890$ & 0.840 & 0.579  & 1.63  & 1.12   & 3.39  & 2.33   \\
          &  & $1.0$   & $-1.5$ & $0.5$ & $4.0$ & $35.6$ &  $42.1$  & $0.171$  & $0.101$   & 0.652 & 0.456  & 1.24  & 0.870  & 2.58  & 1.74   \\
          &  & $-3.0$  & $0.5$  & $1.0$ & $0.0$ & $21.7$ &  $27.2$  & $0.0119$ & $0.00912$ & 0.141 & 0.114  & 0.127 & 0.103  & 0.162 & 0.136  \\
          &  & $3.0$   & $1.5$  & $1.0$ & $2.0$ & $24.7$ &  $29.1$  & $0.376$  & $0.209$   & 0.152 & 0.109  & 0.164 & 0.111  & 0.979 & 0.730  \\
          &  & $-1.0$  & $1.0$  & $0.5$ & $6.0$ & $35.1$ &  $41.1$  & $0.0242$ & $0.00991$ & 0.875 & 0.515  & 1.64  & 1.07   & 3.41  & 2.21   \\
          &  & $-2.0$  & $-0.5$ & $1.0$ & $5.0$ & $20.1$ &  $25.8$  & $0.0221$ & $0.00103$ & 0.243 & 0.142  & 0.191 & 0.113  & 0.270 & 0.157  \\
         &  & \multicolumn{2}{c}{All-sky average} & $1.0$ & $1.0$ & $22.3$ & $26.9$ &  $0.0890$  & $0.0350$& 0.251 & 0.139 & 0.220 & 0.120 & 0.347 & 0.202   \\  \hline\hline
\end{tabular}
\end{center} 
\caption{
Angular resolution for IMBH-IMBH binary coalescence signals from a distance of 100 Mpc 
using the network of three antennas located on the site of TAMA300, LIGO-Hanford and Virgo. 
$\left(S/N \right)_{1}$ denotes 
the total SNR of the network of the three antenna each of which has the only output I.  
$S/N $ denotes 
the total SNR of the three antenna network.  
}
\label{Table:CBC_deltaOmega_multi}
\end{table}

\section{Conclusion}\label{Sec:Conc}
In this paper we have presented a new gravitational-wave (GW) antenna configuration 
with a triple-output TOBA and investigated its performance. 
Previously a single-output TOBA which monitors the angular motions 
caused by GWs only on the horizontal plane had been presented. 
We have developed the antenna configuration by adding two other outputs 
which are given by the angular motions on the vertical planes. 
Thus the gravitational wave signals can be readout from the small rotation of 
the bars on the $x$-$y$, $y$-$z$ and $z$-$x$ planes as can be seen in Fig. \ref{fig:bar}. 

We derive the antenna pattern functions of the triple-output TOBA 
including the effect of the Earth's motion and obtain Eqs. (\ref{Eq:F+}) and (\ref{Eq:Fx}). 
The antenna pattern power is depicted in Fig. \ref{Fig:AntennaPower}, 
which shows the triple-output TOBA has no blind direction 
and is sensitive to GWs from all directions. 
Its detection event rates is better than the single-output configuration 
by a factor of about 1.7. 

We also analyze the accuracies of waveform parameters, focusing on the 
accuracy of the source location on the sky, 
for the two nominal sources, monochromatic sources and binary coalescences. 
For long-duration signals, 
the triple-output TOBA can be treated as the same way as 
the conventional single-output TOBA apart from the improved $S/N$, as expected. 
So the advantage of the triple outputs is merely the accumulation of the $S/N$. 
On the other hand, since the triple-output TOBA 
can break the degeneracy of two polarization modes from short-duration signals even with a single antenna, 
it improves the paramter estimation errors drastically compared with the single-output TOBA. 
Thus, the detection method using a triple-output TOBA we proposed 
would be a powerful tool to search for short-duration signals such as a coalescing binary.  
 
As a future work, it is interesting to investigate potential 
of triple-output TOBAs for testing gravitational theories with GWs. 
Generally, alternative theories of gravity allow GWs 
to have more independent polarization modes up to six \cite{Yunes:2013dva}. 
While several conventional interferometric detectors are required  
to separate a mixture of polarization modes of a GW in detector outputs,  
triple-output TOBAs may be able to put constraints on the non-tensorial modes with less antennas 
because three independent signals can be extracted from each TOBA.

In addition, this work is considered to be applicable not only for the GW detection but also for a prompt earthquake detection \cite{Harms:2014}.
TOBA is sensitive to the gravity gradient caused by the earthquake.
Using two triple-output TOBAs, the origin of the earthquake can be determined.
Therefore, the big earthquake can be alerted much earlier since the earthquake signal travels faster by the gravity gradient than by the seismic motion.

\begin{acknowledgments}
  We thank Jun'ichi Yokoyama for fruitful discussions. 
  One of the authors (KE) also thanks Hirotaka Yuzurihara for helpful comments. 
  This work is supported by Grant-in-Aid for JSPS Fellows Nos. 26$\cdot$8636 (KE) and 24$\cdot$7531 (AS), 
  and Grant-in-Aid for Scientiﬁc Research No. 24244031 (MA). 
\end{acknowledgments}

\end{document}